\documentclass[aps,prb,twocolumn]{revtex4}
\usepackage{graphicx}
\usepackage{dcolumn}
\usepackage{amsmath}
\usepackage{amssymb}

\def\uu{\uparrow\uparrow}
\def\ud{\uparrow\downarrow}

\def\dd{\downarrow\downarrow}
\def\u{\uparrow}
\def\d{\downarrow}
\def\rv{{\bf r}}

\def\ec{\epsilon_c}
\def\ex{\epsilon_x}
\def\exc{\epsilon_{xc}}
\def\gcav{\overline{g}_c}
\begin{document}
\title{Spin resolution of the electron-gas correlation energy: \\
  Positive same-spin contribution}
\author{Paola Gori-Giorgi$^1$ and John P. Perdew$^2$}
\affiliation{$^1$INFM Center for
  Statistical Mechanics and Complexity, and
Dipartimento di Fisica, Universit\`a di Roma ``La Sapienza,'' 
Piazzale A. Moro 2, 00185 Rome, Italy \\ 
$^2$Department of Physics and Quantum Theory Group, Tulane
University, New Orleans, Louisiana 70118 USA}
\date{\today}
\begin{abstract}
The negative correlation energy $\epsilon_c(r_s,\zeta)$ per particle 
of a uniform
electron gas of density parameter $r_s$ and spin polarization $\zeta$ is
well known, but its spin resolution into $\ud$, $\uu$, and $\dd$
contributions is not.  Widely-used estimates are incorrect, and
hamper the development of reliable density functionals and
pair distribution functions.  For the spin resolution,
we present interpolations between high- and low-density limits
that agree with available Quantum Monte Carlo data.  In the low-density
limit for $\zeta = 0$, we find that the same-spin
correlation energy is unexpectedly positive, and we explain why.  We
also estimate the $\u$ and $\d$
contributions to the kinetic energy of correlation.
\end{abstract}
\pacs{71.10.Ca, 71.15.-m, 31.15.Ew, 31.25.Eb}
\maketitle
The uniform electron gas is a paradigm for density functional
theory,\cite{kohnnobel,science,FNM} the most widely-used method 
for electronic structure calculations in both condensed matter physics 
and quantum chemistry.  The
effects of exchange and correlation can be evaluated and
understood in the uniform-density limit, and then transferred
to realistic systems. This is done not only in the local
spin density (LSD) approximation but also beyond LSD in generalized
gradient approximations (GGA's), meta-GGA's, and hybrid functionals.\cite{FNM}
The correlation energy $\epsilon_c(r_s,\zeta)$ per
particle in a uniform gas of density parameter $r_s = (4 \pi n
a_0^3/3)^{-1/3}$ and spin polarization $\zeta = (n_\u-n_\d)/n$ 
(where $n_{\sigma}$ is the density of spin-$\sigma$ electrons and
$n=n_\u+n_\d$) is well
known, for example from Quantum Monte Carlo (QMC) studies\cite{CA,OHB} that
have been parametrized\cite{VWN,PZ,PW} 
to respect known limits, but the
spin resolution of $\epsilon_c$ into $\ud$, $\uu$, and $\dd$ contributions
is not known.  In this work, we determine the spin resolution
for all $r_s$ and $\zeta$ as an interpolation between high- and low-density
limits, consistent with $\zeta=0$ QMC data.\cite{OHB}\par

\begin{figure}[b]
\includegraphics[width=7.5cm]{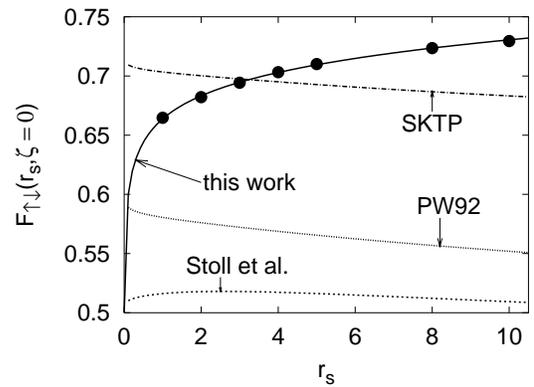} 
\caption{Fraction of $\ud$ correlation energy, $F_{\ud}(r_s,\zeta)=
\ec^{\ud}(r_s,\zeta)/\ec(r_s,\zeta)$ at $\zeta=0$. Our Eq.~(\ref{eq_interp})
is compared with the GSB\cite{GSB} values extracted from 
QMC\cite{OHB} data ($\bullet$),
and with the Stoll {\it et al,}\cite{stoll} PW92\cite{PWgc}, and
SKTP\cite{SKTP} scaling relations. Valence electrons have $2\lesssim r_s
\lesssim 6$.}
\label{fig_stoll}
\end{figure}

This spin resolution is of interest in its own right, and can also be
used in several ways: (i) Some beyond-LSD correlation energy 
functionals need a missing spin resolution\cite{D92} or
have been constructed\cite{becke,scuseria,cohen,altri} on the basis of the 
exchange-like ansatz of Stoll {\it et al,}\cite{stoll}
\begin{equation}
E_c^{\ud}[n_\u,n_\d]\approx E_c[n_\u,n_\d]-E_c[n_\u,0]-E_c[0,n_\d],
\label{eq_stoll}
\end{equation}
for the uniform gas.  This assumption was shown
(using QMC results) to be inaccurate for $\zeta=0$ (see Fig.~\ref{fig_stoll}) 
in Ref.~\onlinecite{GSB}, although the
significance of this observation for density functional theory was not 
fully recognized there. Our work provides a firmer basis than 
Eq.~(\ref{eq_stoll}) for such constructions.  
(ii) Correlation energy functionals such as the
local spin density\cite{kohnnobel} and generalized gradient 
approximations,\cite{PBE}
etc.\cite{jacob} can alternatively be constructed without a 
spin resolution,
but their later spin resolution (to permit
comparison or combination with correlated-wavefunction 
results\cite{cohen,perd,pople})
demands such a resolution for
uniform densities.  (iii) A sophisticated analytic model\cite{GP2} 
is now available
for the pair distribution function\cite{GSB,PWgc,GP2} $g_{xc}(r_s,\zeta,u)$ 
of the uniform gas for all $r_s$ and $\zeta$.  
Our present work provides the missing
ingredient needed to find the corresponding spin-resolved pair
distribution function, which could serve as the starting point for the
development of density functionals such as
spin-resolved weighted density approximations.\cite{AG} (iv) An estimate
can be made for the $\zeta$ dependence of the $\u$ and $\d$ contributions 
to the kinetic energy of correlation, a key ingredient for the approach to spin
dynamics of Qian and Vignale\cite{vignale} and also for the momentum 
distribution\cite{GZ} of a spin-polarized electron gas.\par
%%%%%%%%%%%%%%%%%%%%%%%%%%%%%%%%%%%%%%%%%%%%%%%
%%%%%%%%%%%%%%%%%%%%%%%%%%%%%%%%%%%%%%%%%%%%%%%%
%%%%%%%%%%%%%%%%%%%%%%%%%%%%%%%%%%%%%%%%%%%%%%%
\begin{figure*}
\includegraphics[width=20cm]{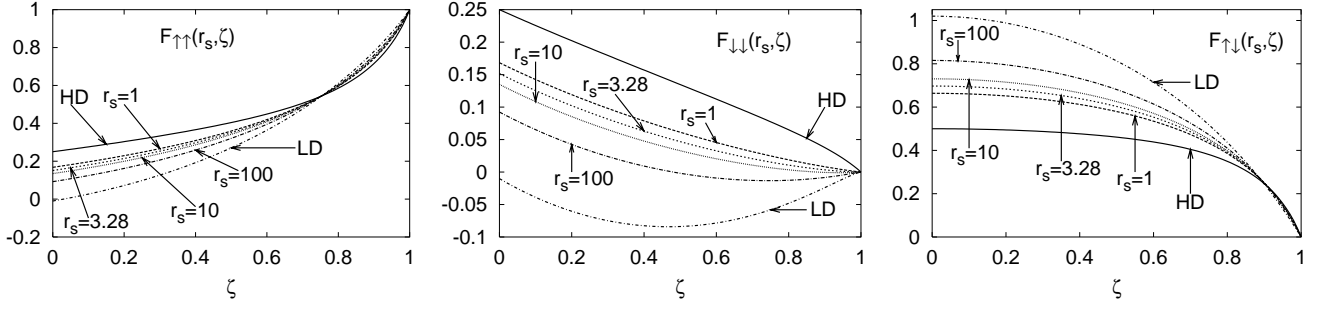} 
\caption{Spin resolution $F_{\sigma\sigma'}(r_s,\zeta)=
\ec^{\sigma\sigma'}(r_s,\zeta)/\ec(r_s,\zeta)$ as a function of $\zeta$ for 
different $r_s$. The high-density (HD) and low-density (LD) limits 
are given in Eqs.~(\ref{eq_HD}) and~(\ref{eq_LDuu}). The $r_s =3.28$ curves
correspond to the SKTP~\cite{SKTP} scaling relation of Eq.~(\ref{eq_SKTP}),
while for other density values ($r_s=1, 10$ and 100) our
interpolation formulas of Eq.~(\ref{eq_interp}) have been used.}
\label{fig_Fss}
\end{figure*}
%%%%%%%%%%%%%%%%%%%%%%%%%%%%%%%%%%%%%%%%%%%%%%%%

We shall first derive exact limits at high densities ($r_s\to 0$)
and extreme low densities ($r_s\to \infty$) using simple physical arguments.
In the latter limit, we find that the same-spin contribution to the
correlation energy can be positive, and we provide an intuitive physical 
picture to explain this feature. 
 While the total correlation energy must be negative, individual terms 
of it (e.g., the kinetic energy of correlation) can be positive.
We then build up and discuss our interpolation
formulas.\par

{\it Definitions} -- 
Correlation effects arise from the Coulomb
interaction, which is a two-body operator. When
evaluating the energy of the system, $\langle \Psi|H|\Psi\rangle$,
one can split the sum over the electron spins into $\ud$, $\uu$, and $\dd$
contributions. The corresponding splitting of the correlation
energy of the uniform electron gas,
\begin{equation}
  \ec(r_s,\zeta)  = \ec^{\ud}(r_s,\zeta)+\ec^{\uu}(r_s,\zeta)
+\ec^{\dd}(r_s,\zeta),
\end{equation}
is the object of this paper. The real-space analysis of the spin-resolved 
correlation energies $\ec^{\sigma\sigma'}(r_s,\zeta)$ is provided by
the correlation holes $n_{\sigma'}\gcav^{\sigma\sigma'}(r_s,\zeta,u)$
(see, e.g., Ref.~\onlinecite{GP2}), where $u=|\rv_1-\rv_2|$ is the
electron-electron distance:
\begin{equation}
\ec^{\sigma\sigma'}(r_s,\zeta)=2\pi\frac{n_{\sigma}}{n}
\int_0^{\infty} n_{\sigma'} \gcav^{\sigma\sigma'}(r_s,\zeta,u)\,u\,du.
\label{eq_real}
\end{equation}
The correlation hole $n_{\sigma'}\gcav^{\sigma\sigma'}(r_s,\zeta,u)$ describes
the change (due to correlation only)
of spin-$\sigma'$ electron density at ${\bf u}$,
when a spin-$\sigma$ electron is at the origin.
$\gcav$ is averaged over coupling strength, while $g_c$ is for full 
coupling strength.
We define fractions $F_{\sigma\sigma'}(r_s,\zeta)$
such that
\begin{equation}
\ec^{\sigma\sigma'}(r_s,\zeta)= \ec(r_s,\zeta)\,F_{\sigma\sigma'}(r_s,\zeta),
\end{equation} 
and we investigate their properties.
In what follows, we use Hartree atomic units, and the 
parametrization of $\ec(r_s,\zeta)$ and its limits 
from Ref.~\onlinecite{PW}.\par

{\it Exact limits} --
When $r_s\to 0$, the Coulomb electron-electron interaction can be treated
as a perturbation to the non-interacting Fermi gas. The first-order
(in the Coulomb potential) correction term gives the exchange energy
$\ex=\ex^{\uu}+\ex^{\dd}$, where
$\ex^{\uu}=-\tfrac{3}{8\pi\alpha r_s}(1+\zeta)^{4/3}$,
$\ex^{\dd}=-\tfrac{3}{8\pi\alpha r_s}(1-\zeta)^{4/3}$,
and $\alpha=(9\pi/4)^{-1/3}$. As for correlation, the real-space
analysis of the exchange energies is provided by the exchange holes
$n_{\sigma}[g_x^{\sigma\sigma}(\zeta,u/r_s)-1]$, which
are analytically known (see, e.g., Ref.~\onlinecite{GP2}).\par
 
The second order correction to the energy of the non-interacting Fermi gas
is the sum of a direct term and
a second-order exchange term. Only the
direct term diverges, and, when a cutoff $\propto 1/\sqrt{r_s}$ (due
to Thomas-Fermi  screening effects) at small wavevectors is introduced,
gives rise to a leading term in $\ec(r_s,\zeta)$, equal to 
$c_0(\zeta)\ln r_s$. The function $c_0(\zeta)$ is exactly 
known.\cite{WP} The direct term 
(Eq.~(5.110) of Ref.~\onlinecite{PN}) can be divided into
$\ud$, $\uu$, and $\dd$ excitation pairs to derive
\begin{equation}
F_{\uu}(r_s\to 0,\zeta)\equiv F_{\uu}^{\rm HD}(\zeta)=
\frac{1+\zeta}{4I(\zeta)}, 
\label{eq_HD}
\end{equation}
with $I(\zeta)=c_0(\zeta)/c_0(0)$, as conjectured 
in Ref.~\onlinecite{WP}. (Since
$F_{\dd}(r_s,\zeta)=F_{\uu}(r_s,-\zeta)$ and
$F_{\ud}=1-F_{\uu}-F_{\dd}$, we
only report formulas for $\uu$.) 
The Stoll {\it et al.} ansatz of Eq.~(\ref{eq_stoll})
is thus correct for $r_s\to 0$ (and for all $r_s$ when
$|\zeta|=1$, but not otherwise).\par

\begin{figure}
\vskip -0.3cm
\includegraphics[width=7.5cm]{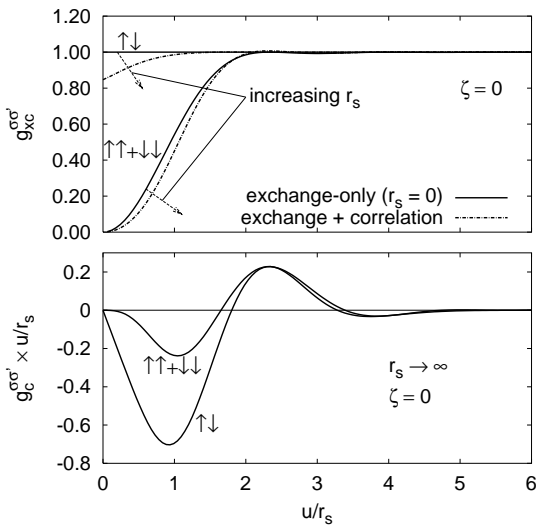} 
\caption{Upper panel: the spin-resolved pair distribution functions
for the paramagnetic gas. The dashed arrows show the trend of the holes
as the coupling strength $r_s$ is increased. 
Lower panel: real-space analysis of the correlation energy in the extreme
low-density limit, for the paramagnetic gas. The results are from the
model of Ref.~\onlinecite{GP2}.}
\label{fig_holes}
\end{figure}

In the opposite or strong-interaction limit, $r_s\to \infty$, 
the long-range Coulomb repulsion 
between the electrons becomes dominant with respect to the kinetic energy, 
and thus with respect to statistics; Coulomb repulsion suppresses
electron-electron overlap so that the electrons no longer know they are
fermions. In this limit, the total 
energy becomes independent\cite{CA,OHB,PW,SPK,GP2} of $\zeta$. Its
leading term in the $r_s\to\infty$ expansion is 
equal to $-d_1/r_s$, where\cite{PW} $d_1\simeq 0.892$, and is
purely potential energy, with no kinetic energy contribution. In this limit,
the total energy is thus equal to the exchange-correlation energy $\exc=
\ex+\ec$. Moreover, since the statistics becomes irrelevant, we expect that
\begin{equation}
\left(\frac{2}{1+\zeta}\right)^2\exc^{\uu}=\left(\frac{2}{1-\zeta}\right)^2
\exc^{\dd}=\frac{2}{(1-\zeta^2)}\exc^{\ud}=\exc,
\label{eq_LD}
\end{equation}
where the prefactors take into account the available numbers of pairs.
In other words, we expect that $\int_0^{\infty}
du \,4 \pi u^2 \overline{g}_{xc}^{\sigma\sigma'}/u$ becomes independent 
of $\sigma$ and $\sigma'$, so that spin structure
becomes unimportant for the exchange-correlation and total energies
(although very important for the correlation energy alone). Then the
$F_{\sigma\sigma'}(r_s\to\infty,\zeta)
\equiv F_{\sigma\sigma'}^{\rm LD}(\zeta)$ are given by
\begin{equation}
F_{\uu}^{\rm LD}(\zeta)=\frac{3(1+\zeta)^{4/3}-2\pi\alpha(1+\zeta)^2\,d_1}
{3[(1+\zeta)^{4/3}+(1-\zeta)^{4/3}]-8\pi\alpha\,d_1}.
\label{eq_LDuu}
\end{equation}
The high- and low-density $F_{\sigma\sigma'}$ are displayed in 
Fig.~\ref{fig_Fss}. We see that, in the spin-unpolarized gas, the 
same-spin ($\uu$+$\dd$)  contribution to
the correlation energy is 50\% when $r_s\to 0$ but roughly 0 when 
$r_s\to\infty$. This can be understood in a simple way. 
The exchange hole seen by the same-spin electrons is deep for 
electron-electron distances $u\lesssim r_s$, as shown in the upper
panel of Fig.~\ref{fig_holes} (solid line, $\uparrow\uparrow+\dd$).
But there is a second length scale, the 
Thomas-Fermi screening length $\sqrt{r_s}$.
For $r_s\to 0$, the important correlations, which determine the
leading term ($\propto \ln r_s$) of $\ec$, arise from this
second length scale,
$\sqrt{r_s}\gg r_s$, and are essentially unaffected by exchange:  the electrons
that participate in this correlation have no way to know if the electron
at $u=0$ is spin-$\u$ or spin-$\d$, so by symmetry the same-spin and
opposite-spin correlation energies are equal.
In the opposite limit, $r_s\to\infty$, 
the antiparallel-spin correlation hole can get deep for $u
\lesssim r_s$, as shown in the upper panel of Fig.~\ref{fig_holes}.

As $r_s$ increases, $g_{xc}^{\ud}$ deviates more and more from
its non-interacting value (equal to 1 for all $u$), the only
constraint being its positiveness. But the 
same-spin correlation hole is "blocked" from doing 
this by the exchange hole (see, again, the upper panel of
Fig.~\ref{fig_holes}).  
Thus the system minimizes its energy by focussing the
correlation on opposite-spin pairs. In the extreme
low-density limit, a simple qualitative picture can be obtained by using
the correlation-hole model of Ref.~\onlinecite{GP2} (in which energetically 
unimportant long-range oscillations are averaged out); in the lower panel of
Fig.~\ref{fig_holes}, we report the corresponding
real-space analysis of $\ec^{\ud}$
and $\ec^{\uu}+\ec^{\dd}$ for $r_s\to\infty$. We see that the 
same-spin correlation hole for $u\lesssim r_s$ cannot get as
deep as the opposite-spin one.\par
%%%%%%%%%%%%%%%%%%%%%%%%%%%%%%%%%%%%%%%%%%%%%%%
%%%%%%%%%%%%%%%%%%%%%%%%%%%%%%%%%%%%%%%%%%%%%%%%

Figure~\ref{fig_Fss} also shows that in the spin-unpolarized gas the
same-spin correlation energy is slightly positive 
($F_{\sigma\sigma}<0$) when $r_s\to\infty$.
In this limit, the electrons correlate strongly, and the exchange-correlation 
holes show a high first-neighbor peak at $u\approx 2 r_s$
(lower panel of Fig.~\ref{fig_holes}).
If the only effect of same-spin correlation were to push same-spin
electrons away from the region of small $u$ and pile them up 
at $u\approx 2 r_s$, then (by the sum rule integral $\int du 4 \pi u^2 
n_{\sigma} \gcav^{\sigma\sigma}(u) = 0$)
the same-spin correlation energy [Eq.~(\ref{eq_real}) with $\sigma=\sigma'$]
would necessarily be negative.  
So, what must really happen is that the same-spin electrons
that accumulate in the peak at $u \approx 2r_s$ include some that have been
pushed out from $u\ll 2r_s$ and some that have been pulled in from $u\gg 2r_s$.
This is again illustrated in the lower panel of Fig.~\ref{fig_holes}.
We interpret the second zero of $g_c$, which appears at large $u$ but only
at large $r_s$, as the energetically important remnant of the long-range
oscillation of $g_c$ in a Wigner crystal. \par

Positive same-spin correlation energy may be an exotic effect, 
but the blockage of negative same-spin correlation also occurs in a
non-magnetic Mott insulator, e.g., an expanded lattice of hydrogen 
atoms where Coulomb correlation
suppresses the $(1s)^2$ configuration on a given site.  The blockage of
same-spin correlation occurs even in a weakly-correlated 
spin-unpolarized system when the correlation hole is spatially constrained,
as for an atom.\cite{cohen,perd,pople}
In the neon atom,
the true (as cited in Ref.~\onlinecite{perd}) anti-parallel-spin correlation 
energy is 65\% of its LSD value,
while the true parallel-spin correlation energy is only 30\%
of its LSD value.

{\it Interpolation between high and low density} -- We want to build up
interpolation formulas for $F_{\sigma\sigma'}(r_s,\zeta)$ that include
all the information available on the spin resolution of $\ec$. Besides
the high- and low-density limits, we have data for 
$F_{\sigma\sigma'}(r_s,0)$, in the range $0.8 \le r_s \le 10$. These
data have been obtained in Ref.~\onlinecite{GSB} 
(GSB) by integrating spin-resolved
QMC correlation holes.\cite{OHB} Moreover,
Schmidt {\it et al.}\cite{SKTP} (SKTP), starting from nearly-exact limits 
of the spin-resolved correlation holes, proposed a 
scaling relation that is
in agreement with the GSB data at $r_s=3.28$, and that,
as shown in Fig.~\ref{fig_Fss} (curves labelled with ``$r_s=3.28$''), 
lies in between the high- and the low-density
limits with a very ``reasonable'' shape. 
The SKTP scaling should thus be a good ``intermediate point'' 
for our interpolation formulas. We thus define
\begin{equation}
F_{\uu}^{\rm SKTP}(\zeta)=\left(\frac{1+\zeta}{2}\right)^{11/6}
\frac{\ec(3.28,1)}{\ec(3.28,\zeta)},
\label{eq_SKTP}
\end{equation}
and we parametrize $F_{\sigma\sigma'}(r_s,\zeta)$ as
\begin{equation}
F_{\sigma\sigma'}(r_s,\zeta)=\frac{F_{\sigma\sigma'}^{\rm HD}(\zeta)+
A_{\sigma\sigma'}(\zeta)\,\sqrt{r_s}+B\,F_{\sigma\sigma'}^{\rm LD}(\zeta)\,r_s}
{1+C\,\sqrt{r_s}+B\,r_s}.
\label{eq_interp}
\end{equation}
$A_{\sigma\sigma'}(\zeta)$ is found by requiring that $F_{\sigma\sigma'}(3.28,
\zeta)=F_{\sigma\sigma'}^{\rm SKTP}(\zeta)$, i.e.,
\begin{eqnarray}
 A_{\sigma\sigma'}(\zeta) & = & \frac{F_{\sigma\sigma'}^{\rm SKTP}(\zeta)-
F_{\sigma\sigma'}^{\rm HD}(\zeta)}{\sqrt{3.28}}+C\,
F_{\sigma\sigma'}^{\rm SKTP}(\zeta)+ \nonumber \\
& & B\,\sqrt{3.28}\,[
F_{\sigma\sigma'}^{\rm SKTP}(\zeta)-F_{\sigma\sigma'}^{\rm LD}(\zeta)].
\end{eqnarray}
The form of Eq.~(\ref{eq_interp}) is motivated by the 
expression for the correlation energy given in Ref.~\onlinecite{PZ}.
The parameters $B$ and $C$ are fixed by a best fit of $F_{\sigma\sigma'}
(r_s,0)$ to the GSB data for $r_s\in [0.8,10]$: $B=0.178488$,
$C=2.856$. 
In Fig.~\ref{fig_stoll}, our $F_{\ud}(r_s,0)$ is compared with the GSB 
data,\cite{GSB} and with the widely-used Stoll 
{\it et al.}\cite{stoll} ansatz of
Eq.~(\ref{eq_stoll}), which 
strongly underestimates the fraction of $\ud$ correlation energy
at metallic and lower densities. The results for the
paramagnetic gas corresponding to
other proposed scaling relations are also shown. 
Our interpolation formulas as functions of $\zeta$, at $r_s=1,\; 10$,
and 100, are displayed in Fig.~\ref{fig_Fss}.\par  
%%%%%%%%%%%%%%%%%%%%%%%%%%%%%%%%%%%%%%%%%%%%%%%%%%%%%%%%%%%%%%%%%%%%%
%\begin{table}
%\begin{ruledtabular}
%\begin{tabular}{lccc}

%$r_s$ & GSB~\cite{GSB} & this work &  Stoll {\it et al.}~\cite{stoll} \\
%\hline
% 1 & 0.665 & 0.664 & 0.516 \\
% 2 &   0.682  &  0.683             & 0.518  \\  
% 5 & 0.710 &  0.709     & 0.516  \\  
% 10 &  0.729   & 0.731   & 0.509 \\ 
%\end{tabular}
%\end{ruledtabular}
%\caption{Fraction of $\ud$ correlation energy, $F_{\ud}(r_s,\zeta)=
%\ec^{\ud}(r_s,\zeta)/\ec(r_s,\zeta)$ at $\zeta=0$. Our Eq.~(\ref{eq_interp})
%is compared with the GSB~\cite{GSB} values extracted from QMC data,
%and with the Stoll {\it et al.}~\cite{stoll} scaling relation.}
%\label{tab_Fud}
%\end{table}
%%%%%%%%%%%%%%%%%%%%%%%%%%%%%%%%%%%%%%%%%%%%%%%%%%%%%%%%%%%%%%%%%%%%%

{\it Kinetic energy of correlation} -- 
Defining\cite{caccamo} $\ec^{\u}=\ec^{\uu}+\tfrac{1}{2}\ec^{\ud}$ 
(with a similar equation for $\d$), the
adiabatic connection
between the non-interacting and interacting limits for
a given density suggests 
estimating the $\u$ and $\d$ contributions (from
the one-particle density matrix) to the kinetic
energy of correlation $t_c=t_c^\u+t_c^\d$ as\cite{caccamo}
\begin{equation}
t_c^{\sigma}(r_s,\zeta)\approx -\frac{\partial}{\partial r_s}\left[
r_s\ec^{\sigma}(r_s,\zeta)\right],
\label{eq_tcd}
\end{equation}
although as Ref.~\onlinecite{ZT} points out there is only one
coupling constant with a Hellmann-Feynman theorem, not one for each
$\sigma$. Taking Eq.~(\ref{eq_tcd}) as a plausible approximation, we find that
the corresponding result for $t_c^\u-t_c^\d$ is in reasonable agreement with
the scaling relation given in Eq.~(29) of Ref.~\onlinecite{vignale}.
(For $r_s\lesssim 5$, the difference 
is less than 3.5\%).
Via Eq.~(\ref{eq_tcd}), we also confirm that, 
for $1\lesssim r_s\lesssim 10$, the quantity 
$(t_c^\u-t_c^\d)/t_c(r_s,\zeta)$
is almost independent of $r_s$, as recently found
in a more sophisticated calculation within the STLS
approximation.\cite{dobson} \par

{\it Conclusions} -- In summary, we have found the spin resolution of 
the electron gas correlation energy, via an approach applied to 
but not restricted to the
three-dimensional uniform electron gas.  Our results can be used to 
understand correlation in
more realistic systems, and to construct improved density functionals
and pair distribution functions.  We have found that the same-spin
correlation energy can be unexpectedly 
but understandably positive.  We have also provided
support for resolutions\cite{vignale,dobson} of the kinetic energy of 
correlation into
$\u$ and $\d$ terms. It is further possible to show that the positive 
spin stiffness of correlation\cite{VWN,PW}
has positive $\ud$ and negative $\uu+\dd$ contributions.
\par

We thank S. De Palo, S. Kuemmel, M. Polini, G. Vignale, J. Tao, 
and P. Ziesche for
useful discussions. Financial support from MIUR 
through COFIN2001 and from the US
National Science Foundation under Grant No. DMR 01-35678 is
acknowledged.
%%%%%%%%%%%%%%%%%%%%%%%%%%%%%%%%%%%%%%%%%%%%%%%%%%%%%%%%%%%%%%%%%%%%%%%

\end{document}